\documentclass{eas}
\usepackage{graphicx}
\usepackage{natbib}
\usepackage{tabularx,ragged2e,booktabs,caption}
\usepackage{ gensymb, textcomp }
\usepackage{hyperref}

\begin{document}

\title{Highlights of the LINEAR survey} 
\author{Lovro Palaversa}\address{Observatoire astronomique de l'Universit\'{e} de Gen\`{e}ve, 51 chemin des Maillettes, CH-1290
Sauverny, Switzerland; \email{lovro.palaversa@unige.ch}}
%
%
\begin{abstract}
Lincoln Near-Earth Asteroid Research asteroid survey (LINEAR) observed
approximately 10,000 deg$^2$ of the northern sky in the period roughly from 1998
to 2013. Long baseline of observations combined with good cadence and depth ($14.5
< r_{SDSS}< 17.5$) provides excellent basis for investigation of variable
and transient objects in this relatively faint and underexplored part of the
sky. Details covering the repurposing of this survey for use in time domain
astronomy, creation of a highly reliable catalogue of approximately 7,200
periodically variable stars (RR Lyrae, eclipsing binaries, SX Phe stars and
LPVs) as well as search for optical signatures of exotic transient events (such
as tidal disruption event candidates), are presented.
\end{abstract}
\maketitle
\section{Introduction}
The MIT Lincoln Laboratory has operated the Lincoln Near-Earth Asteroid Research
(LINEAR) program since 1998 \citep{2000Icar..148...21S}. Unfiltered observations
were performed with two telescopes at the Experimental Test Site located within
the US Army White Sands Missile Range in central New Mexico at an altitude of
1506 m. The program used two essentially identical equatorially mounted, folded
design telescopes with 1m diameter, f/2.5 primary mirrors equipped with
$2560\times1960$ pixel back-illuminated, frame transfer CCD cameras mounted in
the prime focus. Cameras had no spectral filters and in combination with the
telescopes produced $1\degree.60\times1\degree.23$ ($\approx2$ deg$^2$)
field of view with a resolution $2.25^{\prime\prime}$/pix. In 2013, this setup was
replaced by the 3.5m Space Surveillance Telescope.

\begin{figure}[!h]
 \centering
 \includegraphics[scale=0.6]{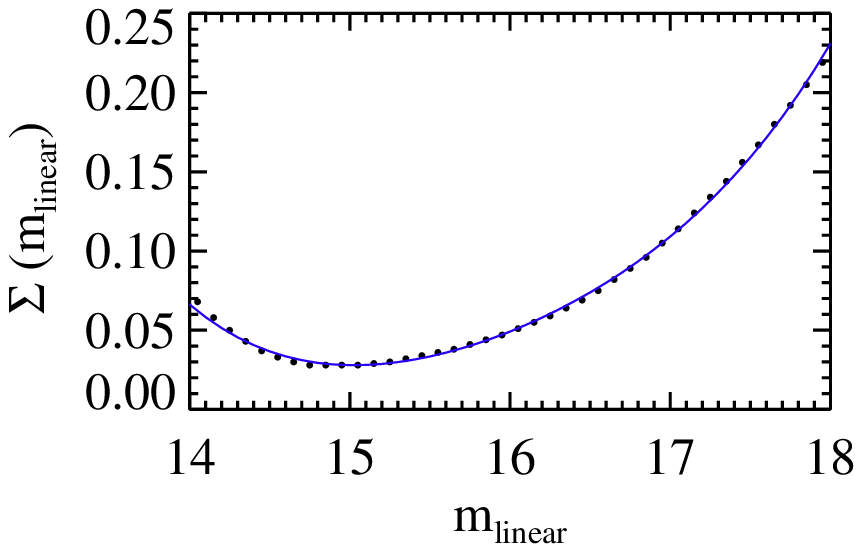}
 \qquad
 \includegraphics[scale=0.6]{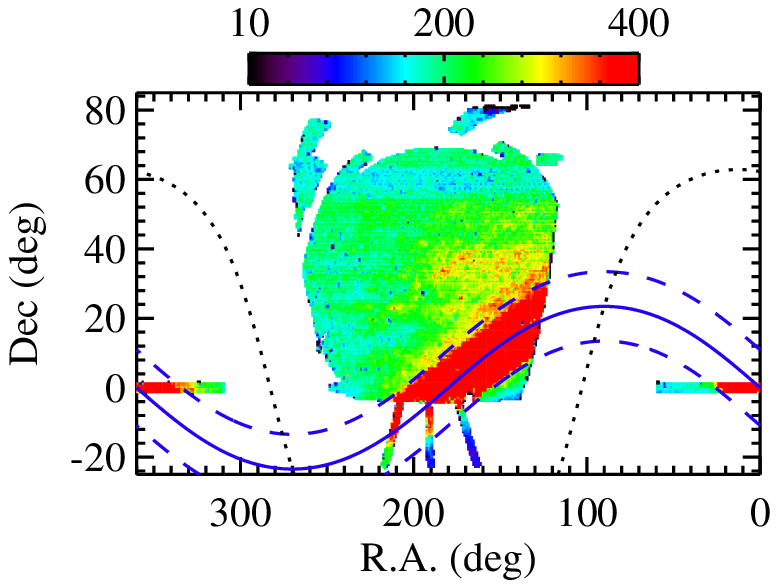}
 \caption{Photometric errors of the recalibrated LINEAR data (left) and average
 number of observations per object (right). Color-coding is according to the
 color bar, with the objects along the ecliptic saturating in red.}
 \label{fig:1}
\end{figure}

First in the series of LINEAR papers (\cite{2011AJ....142..190S}, henceforth
Paper~I) describes the LINEAR survey and photometric recalibration of acquired
data (covering the period from 2002 to 2008), with the SDSS acting as a dense
grid of photometric standards. In the overlapping ~10,000 deg$^2$ of sky between
LINEAR and SDSS, a total of $5\times10^9$ observations of $25\times10^6$ objects
were made. Photometric errors range from ~0.03 mag for sources not limited by
photon statistics to ~0.20 mag at $r_{SDSS}$ = 18 (Figure~\ref{fig:1}).
In this six-year period objects along the ecliptic were observed 400 times on
average, and elsewhere approximately 250 times (Figure \ref{fig:1}). 


\section{The Periodic LINEAR Variables Catalog (PLV)}
Periodic LINEAR variables catalog (PLV) contains 7,194 variable stars classified
with high reliability. Majority of these are new discoveries. The sample is
dominated by RR Lyrae and eclipsing binaries, which account for more than 90\%
of the entire catalog (see Table \ref{tab:1}).

\bigskip
\begin{minipage}{\linewidth}
{\small
\centering
\captionof{table}{PLV catalog} \label{tab:1}
\begin{tabular}{ lcrr}
Type & F [\%] & N \\
\hline \hline
RRAB  & 41  & 2923  \\
RRC  & 14  & 990  \\
EA  & 5  & 357  \\
EB/EW  & 33  & 2385  \\
SXP/DSCT & 2  & 112  \\
LPV  & 1  & 77  \\
Other  & $<5$  & 350  \\
   \hline
Total  & 100  & 7194  \\
\end {tabular}\par
}
\bigskip
\end{minipage}

\begin{figure}
 \centering
 \includegraphics[scale=0.5]{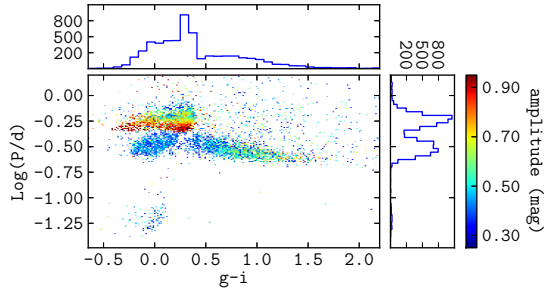}
   \caption{Color-period diagram with color coding according to median
   variability amplitude within a 2D bin. SDSS g-i color on the horizontal and
   logarithm of period in days on vertical axis. Histograms according to the density of
   plotted sources.}
 \label{fig:2}
\end{figure}

In order to create an excedingly pure and complete catalog of variable stars
visual check of all of the variable star candidates had to be performed.
Since the initial $25\times10^6$ object catalog was too large for human
classification, simple statistical cuts were applied in order to bring the
variable star candidate sample to a level manageable on human scale. Application
of cuts on brightness ($14.5 < m_{LINEAR} < 17.5$), likely variability
(${\chi}^2_{dof} > 3$) and variability amplitude ($\sigma > 0.1$ mag) produced a
subsample of 200,000 variable star candidates. The resulting subsample was
phase-folded, crossmatched with 2MASS and WISE catalogs and classified.
Reliability of the classification was checked against AAVSO\footnote{American
Association of Variable Star Observers} Variable Star Catalog (VSX), General
Catalog of Variable Stars (GCVS), SDSS Stripe 82 catalog of variable stars and
CSDR2\footnote{Catalina Surveys Data Release 2} catalog of RR Lyrae. Based on
these comparisons it is estimated that the purity of the catalog is in the 98\%
and completeness in the 55\%-75\% range. Distribution of classified variable stars in color-period
diagram, color coded with amplitude is plotted in Figure \ref{fig:2}.

Further details regarding the creation of the catalog can be found in
\cite{2013AJ....146..101P}(Paper III).
PLV catalog is available online from
\href{http://vizier.cfa.harvard.edu/viz-bin/Cat?J/AJ/146/101}{Vizier}\footnote{\url{http://vizier.cfa.harvard.edu/viz-bin/Cat?J/AJ/146/101}}
 and also from a
 \href{http://www.astro.washington.edu/users/ivezic/linear/PaperIII/PLV.html}{web page}\footnote{\url{http://www.astro.washington.edu/users/ivezic/linear/PaperIII/PLV.html}}
maintained by the authors of Paper III.

In the age of vast astronomical surveys such as Gaia \citep{2012Ap&SS.341..207E}
and LSST \citep{2008arXiv0805.2366I} automated classification becomes
increasingly more important. Particularly, in the case of time-resolved
astronomical phenomena, a need exists for reliable machine learning training
samples that will allow use of automated machine learning methods.

Periodic LINEAR variables catalog (PLV) addresses these needs: it covers a
sufficiently large area of the sky, has no target selection criteria (\ie~it is
unbiased), goes sufficiently deep to overlap with the current (2MASS, GALEX,
SDSS, WISE) and future large surveys (Gaia, LSST) and furthermore has high
reliability and completeness.

With this in mind, several popular automated classification models are described
in the Paper III. Machine learning and data mining python package
\href{http://www.astroml.org}{\mbox{astroML}}\footnote{\url{http://www.astroml.org}}
was extensively used in the process.

\section{Periodic variables and transients from recalibrated LINEAR data} 
PLV catalog contains several interesting variable star subsamples, some of which
have already been subject of research. In particular \cite{2013AJ....146...21S}
used the ab type RR Lyrae stars to explore halo structure and substructure and
confirm the existence of the Oosterhoff dichotomy in the halo.

\begin{figure}[!h]
 \centering
 \includegraphics[scale=0.5]{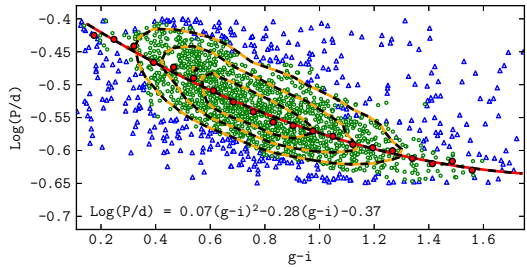}
   \caption{Color-period relation of EW-type binaries. EW type binaries are
   marked with green circles and EB type binaries with blue triangles.}
 \label{fig:3}
\end{figure}

Furthermore Paper III describes 112 candidate SX Phoenicis short period variable
stars outside globular clusters and reproduces the color-period relation for
EW-type eclipsing binaries (Figure~\ref{fig:3}). If SDSS g-i color is taken as
proxy for temperature of these mostly main sequence binaries, relationship
between the decreasing temperature and shortening of the orbital period is
observed. Furthermore, some of the identified short period dM+dM type binary
candidates are found to be below the established 0.2 day period limit.

Long baseline of recalibrated LINEAR data can also be useful in identification
and analysis of transients. One such example is the extreme coronal-line emitter
\mbox{SDSS J095209.56+214313.3} (Palaversa et al., in prep.). Combining the
LINEAR light curve with archived spectroscopic and multi-wavelength observations
allows for precise timing and light curve modelling of this event which occured
in late 2004, while difference imaging constrains the location of the transient
with respect to the host galaxy core.

\section{Conclusion}
Recalibrated LINEAR observations represent a faint and wide time resolved data
set. Well characterized, clean and complete catalogue of LINEAR periodic
variables provides a neccessary machine learning training sample for the large
ongoing and upcoming surveys such as Gaia and LSST. Surveys issuing large
numbers of transient event alerts can also benefit from the classification
provided by the PLV and the complete LINEAR light curve database to which their
more current observations can be compared to. LINEAR data can also be used in
the studies of Galactic structure and pyhsics of binary and eclipsing stars.
Finally, LINEAR data is available on-line via
\href{http://skydot.lanl.gov/}{SkyDOT} and
\href{http://vizier.cfa.harvard.edu/viz-bin/Cat?J/AJ/146/101}{Vizier}.



\end{document}